\documentclass{article}
\usepackage{geometry}
\usepackage{wrapfig}
\usepackage{graphicx}
\usepackage{bm}
\usepackage{amsmath,amssymb}
\usepackage{mathrsfs}
\newcommand{\lw}[1]{\smash{\lower 1.5ex\hbox{#1}}}
\newcommand{\lsim}{\raisebox{-0.6ex}{ $\stackrel{\displaystyle <}{\sim}$ }}
\newcommand{\dint}{\displaystyle\int}

\begin{document}

\title{Analyses of whole transverse momentum distributions in $p\bar p$ and $pp$ collisions by using a modified version of Hagedorn's formula}

\author{Takuya Mizoguchi$^{1}$, Minoru Biyajima$^{2}$, and Naomichi Suzuki$^{3}$\\
{\small $^{1}$National Institute of Technology, Toba College, Toba 517-8501, Japan}\\
{\small $^{2}$Department of Physics, Shinshu University, Matsumoto 390-8621, Japan}\\
{\small $^{3}$Matsumoto University, Matsumoto 390-1295, Japan}}

\date{}
\maketitle

\begin{abstract}
To describe the transverse distribution of charged hadrons at 1.96 TeV observed by the CDF collaboration, we propose a formula with two component, namely, hadron gas distributions and inverse power laws. The data collected at 0.9, 2.76, 7, and 13 TeV by the ALICE, CMS, and ATLAS collaborations are also analyzed using various models including single component models as well as two component models. The results by using modified version of Hagedorn's formula are compared with those by using the two component model proposed by Bylinkin, Rostovtsev and Ryskin (BRR). Moreover, we show that there is an interesting interrelation among our the modified version of Hagedorn's formula, a formula proposed by ATLAS collaboration, and the BRR formula.
\end{abstract}

\section{\label{sec1}Introduction}
First of all, we would like to mention what implications are drawn from the analyses of the transverse momentum spectrum at $\sqrt s = 1.96$ TeV by the CDF collaboration. Authors of the CDF collaboration have reported interesting data on the transverse momentum distributions ($p + \bar p \to$ charged hadrons $+ X$) at $\sqrt s = 1.96$ TeV \cite{Aaltonen:2009ne}. To explain those data, they adopted an inverse power law, given by QCD calculus as follows \cite{Odorico:1982eq,Arnison:1982ed}:
\begin{eqnarray}
  \frac{d^2\sigma}{2\pi p_tdp_t} = A\left(\frac{p_{t0}}{p_{t0}+p_t}\right)^{n_x},
\label{eq_QCD}
\end{eqnarray}
where $A$, $p_{t0}$, and $n_x$ are free parameters that are estimated using experimental data (for further information, see useful reviews in Refs. \cite{Kondou:1980aa,Uematsu:1980aa,Green:2005aa,Rak:2013aa}). The unit of the coefficient $A$ is [mb/GeV$^2$].

\begin{figure}
  \centering
  \includegraphics[width=0.48\columnwidth]{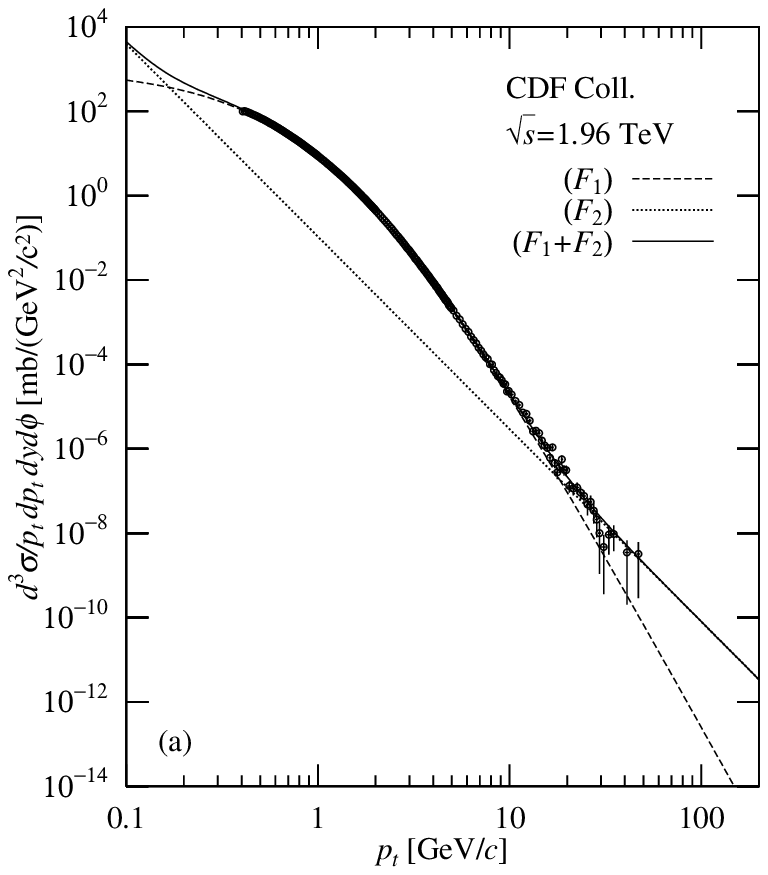}
  \includegraphics[width=0.48\columnwidth]{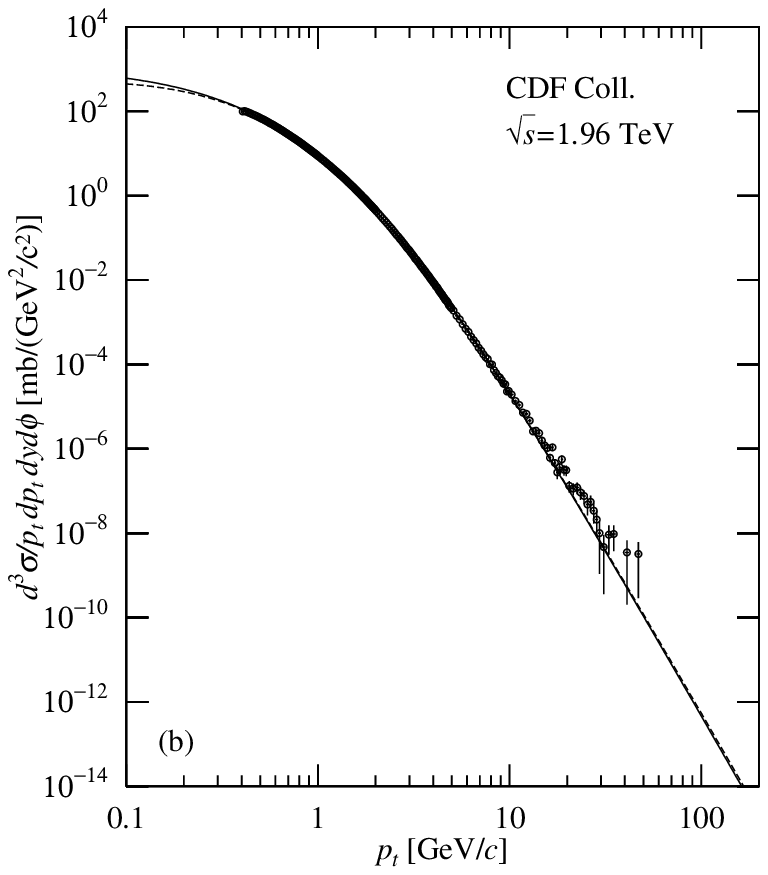}
  \caption{\label{fig_CDF}(a) Analysis of $p_t$ distribution at $\sqrt s = 1.96$ TeV using Eq. (\ref{eq_CDF}). $F_1$ and $F_2$ represent the results of the first and second term of Eq. (\ref{eq_CDF}), respectively, whereas ($F_1 + F_2$) is the sum of these terms. (b) Analysis of $p_t$ distribution at $\sqrt s = 1.96$ TeV using Eq. (\ref{eq_QCD}) (solid line) and Eq. (\ref{eq_NEXT}) (dashed line). The experimental data points are given along with their statistical and systematic errors.}
\end{figure}

As can be seen by the results in Table \ref{tab_CDF}, Eq.~(\ref{eq_QCD}) cannot explain the data. Thus, for charged particle production, a second term was introduced by the CDF collaboration as follows:
\begin{eqnarray}
  \frac{d^2\sigma}{2\pi p_tdp_t} = A\left(\frac{p_{t0}}{p_{t0}+p_t}\right)^{n_x} + B\left(\frac 1{p_t}\right)^{n_s},
\label{eq_CDF}
\end{eqnarray}
where $B$ and $n_s$ are free parameters. Our reanalyses of the data are shown in Table \ref{tab_CDF} and Fig. \ref{fig_CDF}(a). The unit of the second coefficient $B$ is [mb$\cdot$GeV$^{(n_s-2)}$]. Indeed, better values for $\chi^2$than those provided by Eq.~(\ref{eq_QCD}) are possible. Here we present our reanalyses using the statistical and systematic errors presented at Durham HepData \cite{HepData:2009aa}.

Herein, it is worthwhile to investigate the role of the single component model, i.e., the non-extensive approach \cite{Biyajima:2004ub,Wong:2015mba,Barnafoldi:2011zz,Cleymans:2012ac}. The formula for this is as follows:
\begin{eqnarray}
  \frac{d^2\sigma}{2\pi p_tdp_t} = Ae_q^{-\beta x_{\pi}} = \frac A{(1+\beta (q-1)x_{\pi})^{1/(q-1)}},
  \label{eq_NEXT}
\end{eqnarray}
where $x_{\pi} =\sqrt{m_{\pi}^2 + p_t^2}$ and $\beta =1/k_BT$. $A$, $T$, and $(q-1)$ are free parameters, and $(q-1)$ is named as the Tsallis parameter. $n=1/(q-1)$ is sometimes used. The unit of the coefficient $A$ is [mb/GeV$^2$]. It should be stressed that Eq. (\ref{eq_NEXT}) contains the following interesting properties:
\begin{eqnarray*}
\mbox{Eq. }(\ref{eq_NEXT}) \sim \left\{
\begin{array}{l}
e^{-\beta x_{\pi}}\quad \mbox{for } p_t\to 0\\
(\beta (q-1)x_{\pi})^{-1/(q-1)}\quad \mbox{for } p_t\to \infty.
\end{array}
\right.
\end{eqnarray*}
Indeed the second row term reproduces the inverse power law. The results of our analyses of the data using Eq. (\ref{eq_NEXT}) are presented in Table \ref{tab_CDF} and Fig. \ref{fig_CDF}(b). The values of $\chi^2$ in Table \ref{tab_CDF} demonstrate that the role of Eq. (\ref{eq_NEXT}) is similar to that of Eq. (\ref{eq_QCD}). 

Moreover, the ATLAS collaboration proposed the following formula \cite{Aad:2010ac}:
\begin{eqnarray}
    \frac{d^2\sigma}{2\pi p_tdp_t} = A \left.\tanh^{-1}\left(\dfrac{p_t\sinh \eta}{\sqrt{m_{\pi}^2+p_t^2\cosh^2 \eta}}\right)\right|_{\eta = \eta_c} \times \frac{(n-1)(n-2)}{(nT+m_{\pi}(n-1))(nT+m_{\pi})}\left(\frac{nT+x_{\pi}}{nT+m_{\pi}}\right)^{-n}.
  \label{eq_ATLAS}
\end{eqnarray}
where $\eta_c$ is the centrality of pseudorapidity $|\eta|<\eta_c$, $n=1/(q-1)$ is the parameter introduced in Eq. (\ref{eq_NEXT}). The second function $f(n,\;T,\;p_t,\;m_{\pi})$ in Eq. (\ref{eq_ATLAS}) is the normalized Tsallis distribution whose role is the same as Eq. (\ref{eq_NEXT}). Our results (obtained using Eq. (\ref{eq_ATLAS})) are also given in Table \ref{tab_CDF}. The unit of the coefficient $A$ is [mb], because that of the denominator is [GeV$^{-2}$].

\begin{table*}[htbp]
  \centering
  \caption{\label{tab_CDF}Estimated values from the analysis of $p_t$ distribution at $\sqrt s = 1.96$ TeV using Eqs. (\ref{eq_QCD}), (\ref{eq_CDF}), (\ref{eq_NEXT}) and (\ref{eq_ATLAS}) ($\eta_{\rm c}=1.0$). The statistical and systematic errors are given along with the data.}
  \begin{tabular}{ccccccc}
  \hline
  Eq. & $\chi^2/$ndf & $A$ & $p_{t0}$ (GeV) & $n_x$ & $B$ & $n_s$ \\
  \hline
  (\ref{eq_QCD}) & 221/227 & 1160$\pm$20 & 1.18$\pm$0.01 & 7.97$\pm$0.02 
& --- & --- \\
  (\ref{eq_CDF}) & 131/225 & 1010$\pm$30 & 1.28$\pm$0.02 & 8.22$\pm$0.05 
& 0.107$\pm$0.034 & 4.56$\pm$0.09 \\
  \hline
  \hline
  & $\chi^2/$ndf & $A$ & $T$ (GeV) & $q-1$ \\
  \hline
  (\ref{eq_NEXT})  & 198/227 & 1360$\pm$30 & 0.144$\pm$0.001 & 0.1269$\pm$0.0005\\
  (\ref{eq_ATLAS}) & 193/227 & 1430$\pm$27 & 0.142$\pm$0.001 & 0.1271$\pm$0.0004\\
  \hline
  \end{tabular}
\end{table*}

The $p_t$ distribution at 1.96 TeV obtained by the CDF collaboration suggests that a two component model is necessary to explain the distribution. These results further suggest that the second term provides a reasonable way of explaining the transverse momentum spectrum. Moreover, they demonstrate that we must investigate whether the second term of Eq. (\ref{eq_CDF}) provides a unique solution.

In section \ref{sec2}, we investigate several formulas based on the two component models. In section \ref{sec3}, we present the analyses of other $p_t$ spectra obtained by the ALICE, CMS, and ATLAS collaborations \cite{Abelev:2013ala,Adam:2015pza,Chatrchyan:2011av,Aad:2010ac}. In section \ref{sec4}, concluding remarks and discussion are presented.

\section{\label{sec2} Analysis of the data by CDF Collaboration using two component models}
\begin{figure}
  \centering
  \includegraphics[width=0.48\columnwidth]{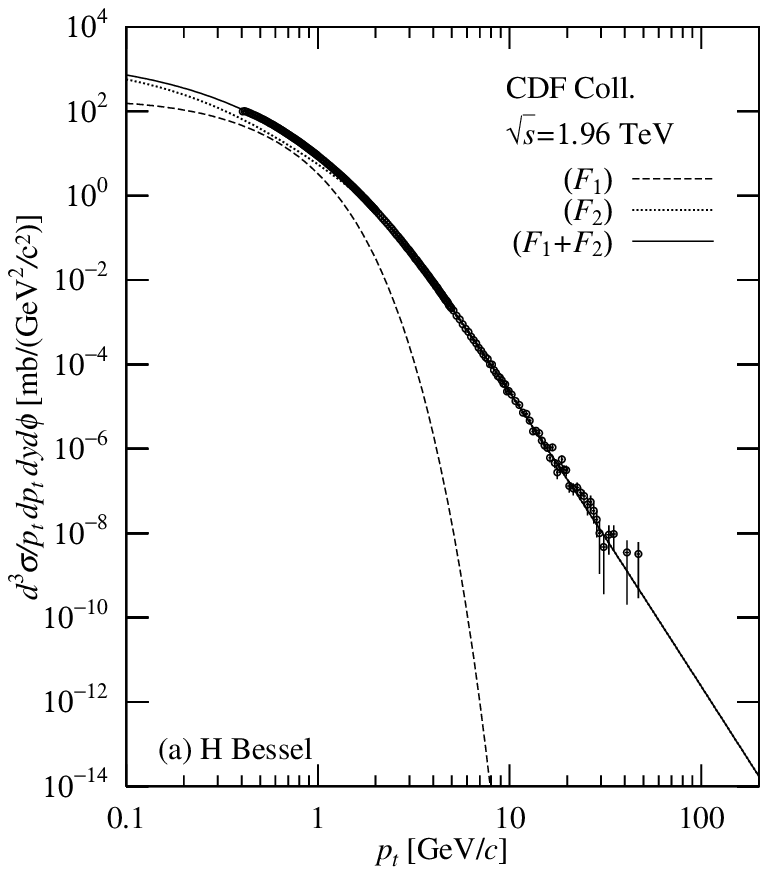}
  \includegraphics[width=0.48\columnwidth]{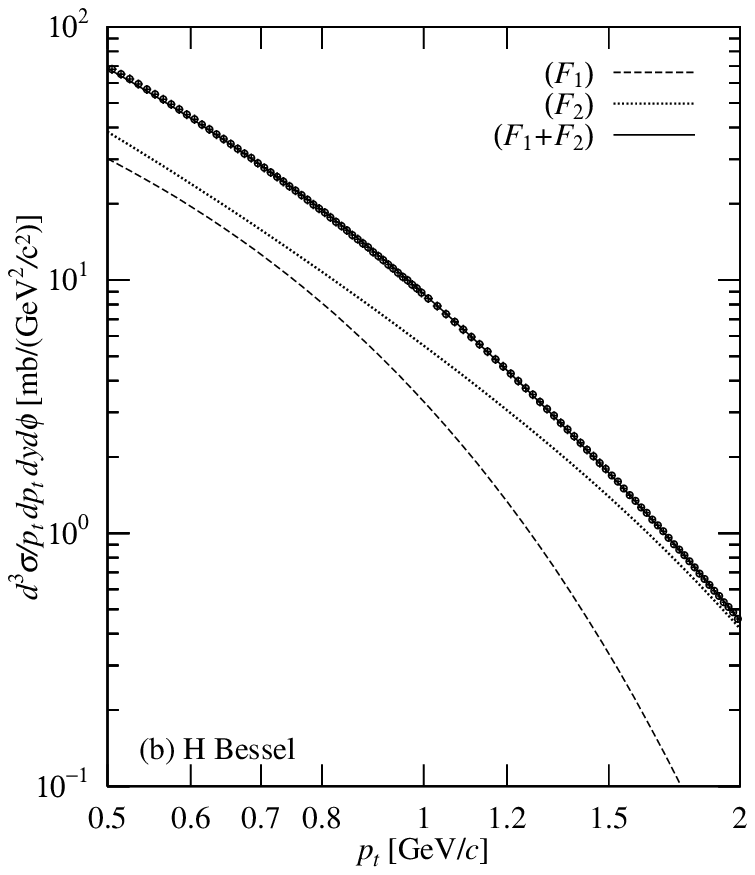}
  \caption{\label{fig_Hag}(a) Analysis of $p_t$ distribution at $\sqrt s = 1.96$ TeV using Eq. (\ref{eq_Hag}), which is the Hagedorn's Bessel formula. $F_1$ and $F_2$ represent the results using the first and second terms of Eq. (\ref{eq_Hag}), respectively. ($F_1 + F_2$) is the sum of these terms. (b) Enlarged section of panel (a). Note that the first term always yields lower values than the second term.}
\end{figure}

\paragraph{2-I)} In 1983, Hagedorn proposed the following formula, which includes the Planck distribution of a hadron gas ($p_t \le 1 - 2$ GeV) \cite{Hagedorn:1983wk,Biyajima:2006mv} and an inverse power law distribution inspired by QCD calculus \cite{Odorico:1982eq,Arnison:1982ed}. The second term in the equation is characterized by a step function defined by $\theta = 1$ ($p_t > p_{t1} = 1.0 - 2.0$ GeV), and $\theta = 0$ ($p_t < p_{t1}$).
\begin{eqnarray}
\frac{d^2\sigma}{2\pi p_t dp_t} = A_p x_{\pi}\sum_{l=1}^N K_1(l\beta x_{\pi}) 
 +A_h\theta(p_t,\: p_{t1})\left(\frac{p_{t0}}{p_{t0}+p_T}\right)^{n_x},
\label{eq_Hag}
\end{eqnarray}
where $K_1$ is the modified Bessel function of the second type with $\beta = 1/T$. The unit of the first coefficient $A_p$ is [mb/GeV$^3$]. That of the second one $A_h$ is [mb/GeV$^2$]. The former is obtained by integrating the rapidity $y$ with the limits $-\infty < y < \infty$. For concrete analyses, it is very difficult to adopt a sharp step function. Hereafter, the step function $\theta$ is assumed to be of the form of a Fermi-Dirac (FD) distribution with inverse temperature $\beta_{\rm FD}$ \cite{Mizoguchi:2008ns}:
\begin{eqnarray}
\theta(\beta_{\rm FD},\: p_t,\: p_{t1}) = \frac 1{\exp[-\beta_{\rm FD} (p_t-p_{t1})]+1}. 
\label{eq_Step}
\end{eqnarray}
Our analyses results of the data using Eq. (\ref{eq_Hag}) with Eq. (\ref{eq_Step}) are presented in Table \ref{tab_Hag} and Fig. \ref{fig_Hag}.
\begin{table*}[htbp]
  \centering
  \caption{\label{tab_Hag}Estimated values for the analysis of $p_t$ distribution at $\sqrt s = 1.96$ TeV using Eq. (\ref{eq_Hag}) with $N=5$. The unit of $A_p$ is [mb/GeV$^3$] and that of $A_h$ is [mb/GeV$^2$].}
  \begin{tabular}{cccccccc}
  \hline
  $\chi^2/$ndf & $A_p$ & $T$ & $A_h$ & $p_{t0}$ (GeV) & $n$ & $p_{t1}$ (GeV) & $\!\beta_{\rm FD}$ (GeV$^{-1}\!$)\\
  \hline
  $\!$77.5/223$\!$ & $\!$823$\pm$96$\!$ & $\!$0.201$\pm$0.003$\!$ & $\!$34400$\pm$1100$\!$ 
& $\!$0.536$\pm$0.003$\!$ & $\!$7.11$\pm$0.01$\!$ & $\!$1.43$\pm$0.03$\!$ & 2.12$\pm$0.06\\
  \hline
  \end{tabular}
\end{table*}

As seen in Table \ref{tab_Hag}, the coefficient $A_h$ is much larger than $A_p\sqrt{s_0}$ ($A_h \gg A_p\sqrt{s_0}$ with $\sqrt{s_0}=1.0$ GeV, which is a typical value for the energy). In other words, the contribution of the second term, which is based on QCD calculus, is large even in the region in which $p_t < 1 - 2$ GeV. This behavior appears to be contradictory to the assumption made by Hagedorn (Fig. \ref{fig_Hag}).

\paragraph{2-II)} Compared Eq. (\ref{eq_Hag}) with Eq. (\ref{eq_NEXT}), we assume the Bose--Einstein distribution for $Ed\sigma/d^3p = d\sigma/2\pi p_t dp_t dy$. Moreover, taking into account the empirical condition $|\eta| \le 1$ discovered by the CDF collaboration, we propose here the following formula based on \cite{Biyajima:2003nv}, which includes the power law of the dipole-like expression proposed by the CFS collaboration in \cite{Innes:1977ae} (see \cite{Kondou:1980aa}) and the QCD-inspired term already proposed in a different point of view in \cite{Michael:1976pz}:
\begin{eqnarray}
\frac{d^2\sigma}{2\pi p_tdp_t} 
 &=& A_p^{(M)}\int_{-\eta_{\rm max}}^{\eta_{\rm max}}d\eta \frac{J(\eta)}{\exp(\beta [x_{\pi}\cosh y(\eta)])-1}
  + A_h\theta(\beta_{\rm FD},\: p_t,\: p_{t1})\left(\frac{p_{t0}^2}{p_{t0}^2+p_t^2}\right)^{n_x},
 \nonumber\medskip\\
 &\equiv& F_1(\mbox{hadron gas})+F_2(\mbox{QCD inspired term})
  \label{eq_BED}
\end{eqnarray}
where 
\begin{eqnarray}
y(\eta) &=& \frac 12 \ln \frac{\sqrt{1+(m_{\pi}/p_t)^2+\sinh^2 \eta}+\sinh \eta}{\sqrt{1+(m_{\pi}/p_t)^2+\sinh^2 \eta}-\sinh \eta},
 \label{eq_rap}\\
J(\eta) &=& \frac{\cosh \eta}{\sqrt{1+(m_{\pi}/p_t)^2+\sinh^2 \eta}},
 \label{eq_Jac}
\end{eqnarray}
and where $J$ is a Jacobian. The unit of $A_p^{(M)}$ and $A_h$ is [mb/GeV$^2$].

In this description, there are various combinations between the two sets $\{p_{t1}(i):\,i=1,\,2,\,\cdots \}$ and $\{\beta_{\rm FD}(i):\,i=1,\,2,\,\cdots \}$ in Eq. (\ref{eq_BED}) for the MINUIT program. To take into account the requirement for the step function according to Hagedorn's equation ($\theta(x) =0\:(x>0)$ and $=0\:(x<0)$), we have to introduce the following realistic constraints:
\begin{eqnarray}
\left\{
\begin{array}{l}
\dfrac{F_1}{F_2} \lsim 0.01(2)\quad \mbox{at}\ p_t = 0.2\ \mbox{GeV}, \medskip\\
\dfrac{F_2}{F_1} \lsim 0.01(2)\quad \mbox{at}\ p_t = 2.0\sim 3.0\ \mbox{GeV}
\end{array}
\right.
\label{eq_con}
\end{eqnarray}
The contributions of the two components can be reasonably calculated using Eq. (\ref{eq_con}), which is desirable because the change in a rigorous step function is very steep for it to be adoptable.

\begin{figure}
  \centering
  \includegraphics[width=0.48\columnwidth]{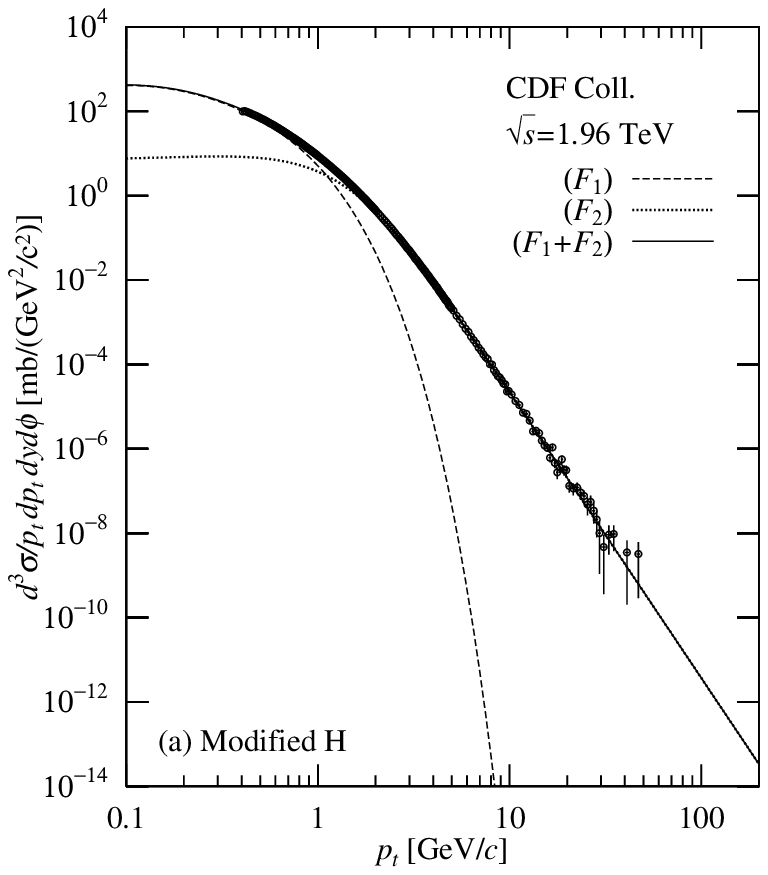}
  \includegraphics[width=0.48\columnwidth]{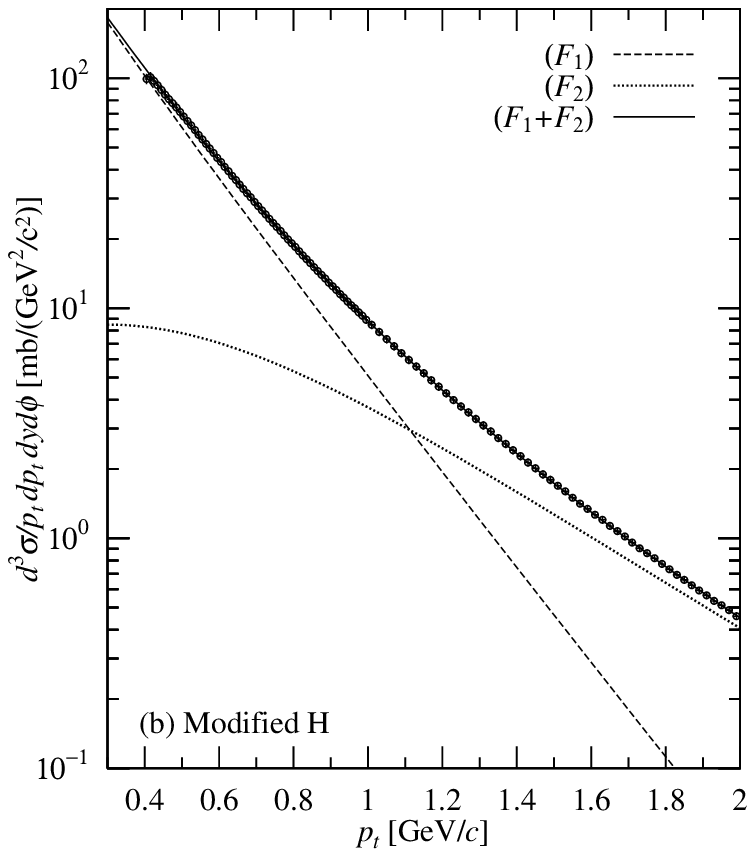}\\
  \includegraphics[width=0.48\columnwidth]{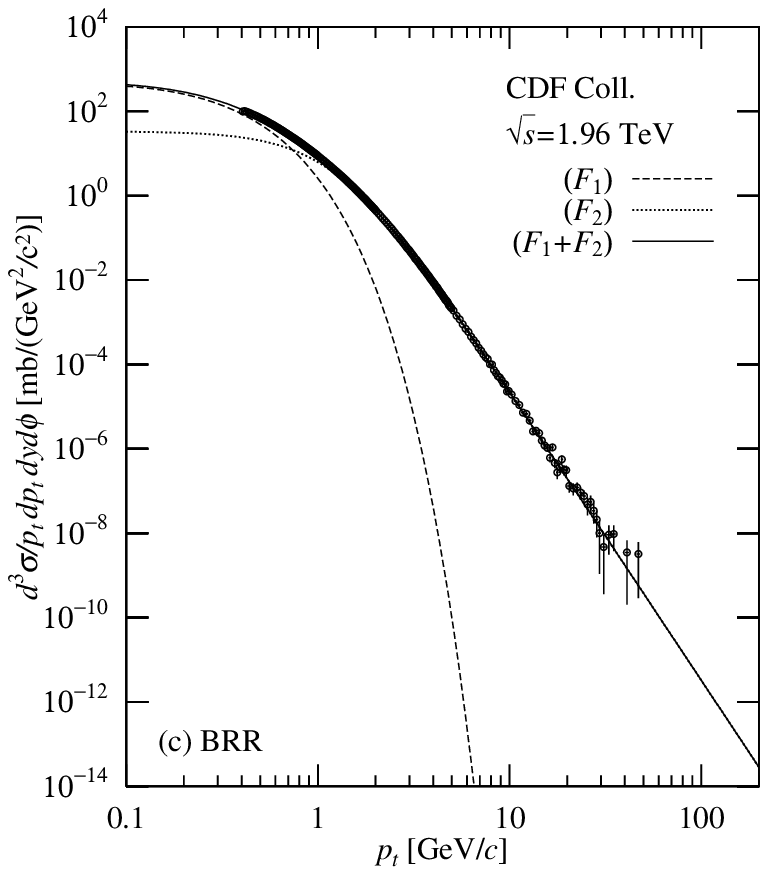}
  \includegraphics[width=0.48\columnwidth]{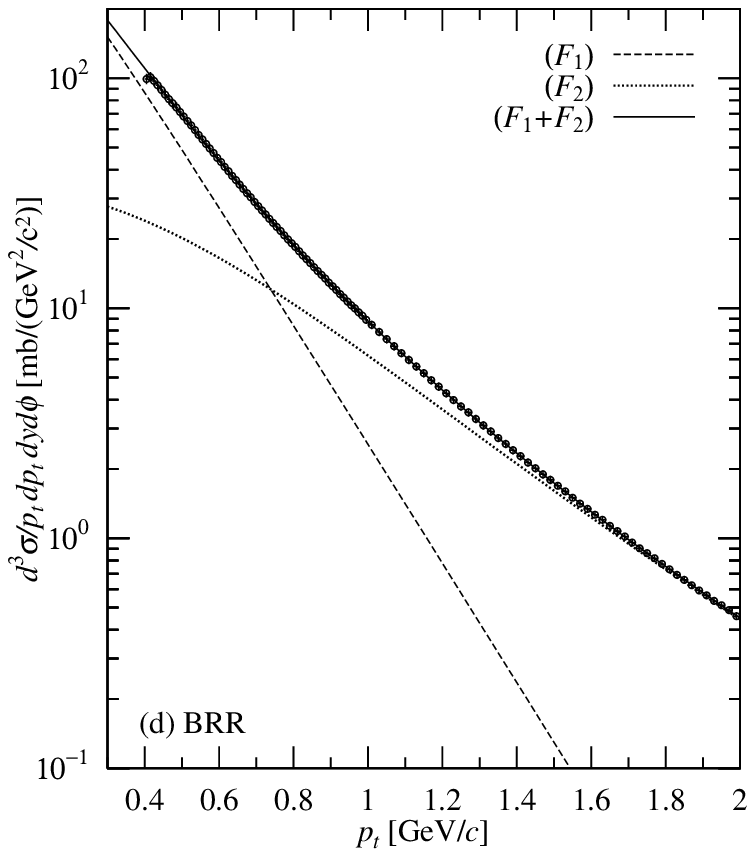}
  \caption{\label{fig_BED}(a) Analysis of $p_t$ distribution at $\sqrt s = 1.96$ TeV using Eq. (\ref{eq_BED}), which is referred to as the ``Modified Hagedorn's formula.'' $F_1$ and $F_2$ show the results of the first and second terms of Eq. (\ref{eq_BED}), respectively. ($F_1 + F_2$) is the sum of these terms. (b) Enlarged section of panel (a). Panels (c) and (d) show the analysis of $p_t$ distribution at $\sqrt s = 1.96$ TeV using Eq. (\ref{eq_Byl}). The notation used in panels (c) and (d) is the same as that in panels (a) and (b).}
\end{figure}

Our analyses results of the data using Eq. (\ref{eq_BED}) and the $\theta$ function of Eq. (\ref{eq_Step}) are presented in Table \ref{tab_BED} and Fig. \ref{fig_BED}. As seen in Fig. \ref{fig_BED}, the second term is smaller than the first term in the region in which $p_t<1$ GeV. The crossing occurs at approximately 0.9 GeV. This behavior satisfies Hagedorn's assumption, and the $\theta$ function works appropriately.
\begin{table}[htbp]
  \centering
  \caption{\label{tab_BED}Estimated values for the analysis of $p_t$ distribution at $\sqrt s = 1.96$ TeV using Eqs. (\ref{eq_BED}) and (\ref{eq_Byl}) with $\varepsilon=1$. The unit of $A_p^{(M)}$, $A_e$, and $A_h$ is [mb/GeV$^2$]. That of the crossing is [GeV].}
  \begin{tabular}{cccccccccc}
  \hline
  Eq. & $\chi^2/$ndf & $A_p^{(M)}$, $A_e$ & $T$, $T_e$ (GeV)   & $A_h$ & $\!\!\! p_{t0}$, $T_{\tiny BRR}$(GeV)$\!\!$ & $n_x$, $n$ & $\!\!$crossing$\!\!$ & $p_{t1}$(GeV)$\!\!$ & $\!\!\beta_{\rm FD}$ (GeV$^{-1}$)\\
  \hline
  (\ref{eq_BED})       & $\!\!$70.7/223 & 406$\pm$76   & 0.22$\pm$0.02 & 61$\pm$42 & 1.1$\pm$0.1  & 3.4 & 1.1 & 1.14$\pm$0.22$\!\!$   & 1.84$\pm$0.35\\
  (\ref{eq_Byl})$\!\!$ & $\!\!$71.6/225 & 477$\pm$19   & 0.17$\pm$0.00 & 33$\pm$2 & 0.68$\pm$0.01 & 3.4 & 0.7 & --- & ---\\
  \hline
  \end{tabular}
\end{table}

\paragraph{2-III)} Recently, Bylinkin, Rostovtsev, and Ryskin \cite{Bylinkin:2014qea,Bylinkin:2012bz,Bylinkin:2015xya} proposed the following interesting formula that includes the Maxwell--Boltzmann distribution that describes pion gas ($p_t < 1 - 2$ GeV) and the kappa distribution adopted in plasma physics \cite{Vasyliunas:1968aa,Hasegawa:1985aa}.
\begin{eqnarray}
\frac{d^2\sigma}{2\pi p_tdp_t} = A_e \exp(-E_{tkin}/T_e) + \frac{A_h}{(1+p_t^2/(T_{\tiny BRR}^2n))^n},
\label{eq_Byl}
\end{eqnarray}
where $E_{tkin} = \sqrt{m_{\pi}^2 + p_t^2}-m_{\pi}$. $A_e$, $T_e$, $A$, $T_{\tiny BRR}$, and $n$ are free parameters. The subscript ``BRR'' refers to the author's names ``Bylinkin, Rostovtsev, and Ryskin.'' It can be noted that the subtracted mass term in $E_{tkin}$ corresponds to the role of the chemical potential $\mu$ in Eq. (\ref{eq_BED}), provided that ``$\beta x_{\pi}\cosh y(\eta) - \mu$'' is assumed in Eq. (\ref{eq_BED}). Our analyses results of the data using Eq. (\ref{eq_Byl}) are presented in Table \ref{tab_BED} and Fig. \ref{fig_BED}.

As seen in Fig. \ref{fig_BED}(c) and (d), the crossing occurs at approximately 0.7 GeV without the step function $\theta$. The hadron (pion) gas mainly contributes in the region in which $p_t < 1 - 2$ GeV, in contrast to case 2-I). The second term reproduces the entire distribution above $p_t>2 - 3$ GeV.

Conversely, an interesting relation between the results obtained using Eqs. (\ref{eq_BED}) and (\ref{eq_Byl}) can be observed. That is expressed as follows:
\begin{eqnarray}
p_{t0}^2 \cong T_{\tiny BRR}^2\cdot n
\label{eq_pt0}
\end{eqnarray}
This relation means that the roles of the second terms in Eqs. (\ref{eq_BED}) and (\ref{eq_Byl}) are almost the same.

\section{\label{sec3} Analyses of the data at $\sqrt s = 2.76$, 7 , and 13 TeV by the ALICE, CMS, and ATLAS collaborations in terms of Eqs. (\ref{eq_NEXT}), (\ref{eq_ATLAS}), (\ref{eq_BED}), and (\ref{eq_Byl})}
In this study, we have analyzed the data at $\sqrt s = 2.76$, 7 , and 13 TeV obtained by the ALICE, CMS, and ATLAS collaborations \cite{Abelev:2013ala,Adam:2015pza,Chatrchyan:2011av,Aad:2010ac}; these data are available in the region in which $p_t \le 1.0$ GeV. It is necessary to investigate whether our formula would be useful in the analyses of data at different energies.

\begin{figure}
  \centering
  \includegraphics[width=0.48\columnwidth]{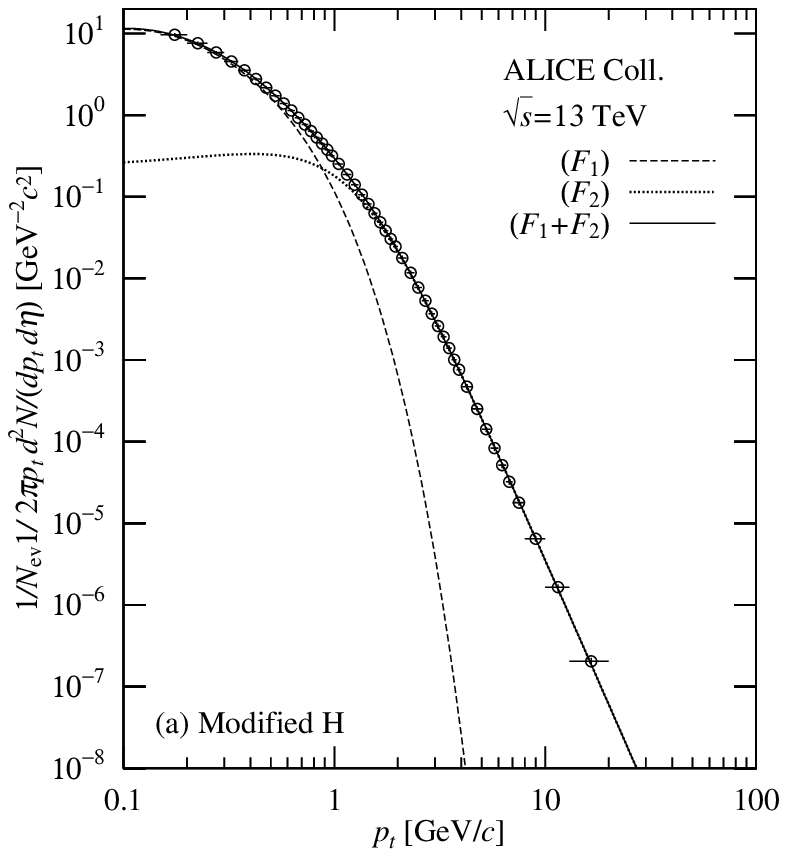}
  \includegraphics[width=0.48\columnwidth]{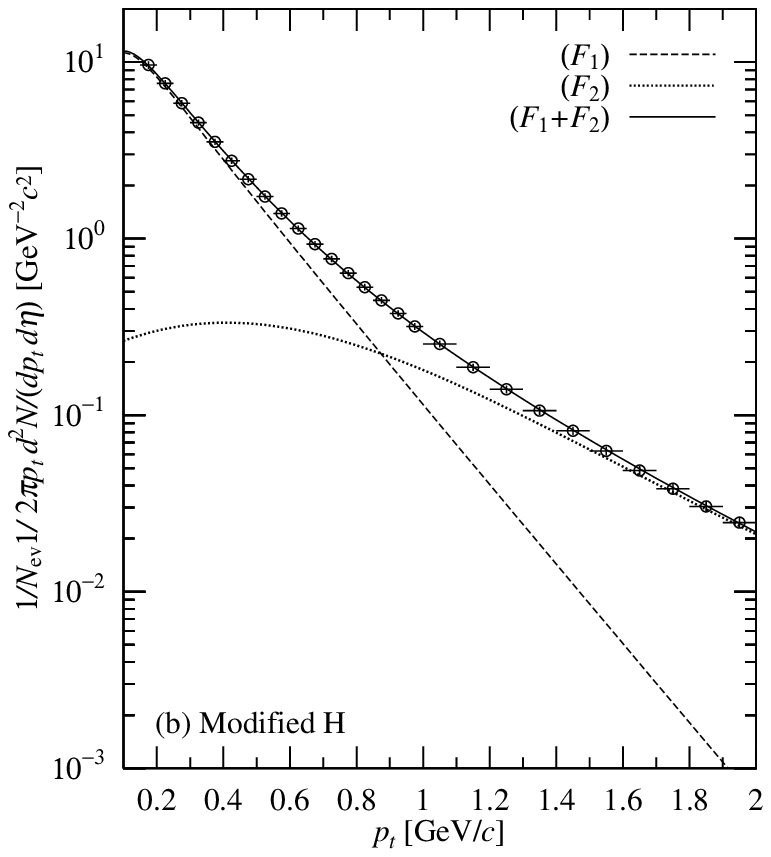}
  \caption{\label{fig_ALICE}(a) Analysis of $p_t$ distribution at $\sqrt s = 13$ TeV using Eq. (\ref{eq_BED}). $F_1$ and $F_2$ represent the results of the first and second terms of Eq. (\ref{eq_BED}), respectively. ($F_1 + F_2$) is the sum of these terms. (b) Enlarged section of panel (a).}
\end{figure}

Our analyses results using Eqs. (\ref{eq_NEXT}), (\ref{eq_ATLAS}), and (\ref{eq_BED}) are presented in Table \ref{tab_7TeV} and Fig. \ref{fig_ALICE}. For comparisons, the results using Eq. (\ref{eq_Byl}) are also presented in Table \ref{tab_7TeV}. Therein, values of $\chi^2$'s by the single component models are larger than those by two component models. This is not clear at present whether or not we need an additional term in Eqs. (\ref{eq_NEXT}) and (\ref{eq_ATLAS}). 

As seen in Fig. \ref{fig_ALICE}(b), our analysis of data at $\sqrt s = 13$ TeV by ALICE collaboration shows that the crossing occurs at 0.9 GeV. Therefore, the second term, which is based on QCD calculus, decreases in the region in which $p_t < 1.0$ GeV. 

\begin{table*}[htbp]
  \caption{\label{tab_7TeV}Estimated values for the analyses of $p_t$ distributions at $\sqrt s =$ 2.76, 7, and 13 TeV using Eqs. (\ref{eq_NEXT}), (\ref{eq_ATLAS}), (\ref{eq_BED}), and (\ref{eq_Byl}) with $\eta_{\rm c} =0.8$ (ALICE), 2.4 (CMS), and 2.5 (ATLAS). Numerical factors in parentheses refer to the collision energies (TeV). The unit of the coefficient $A$ is [mb/GeV$^2$]. That of $A_p^{(M)}$, $A_e$, and $A_h$ is [mb/GeV$^2$]. That of the crossing is [GeV].}
  \begin{tabular}{lccccccccc}
  \hline
Eq. (\ref{eq_NEXT}) & $\chi^2/$ndf & $A$ & $T$ (GeV) & $q-1$\\
\hline
ALICE(2.76)& 38.9/57 & 1670$\pm$70  & 0.13 & $\!\!$0.14\\
ALICE(7) & 68.8/62   & 2190$\pm$90  & 0.13 & $\!\!$0.15\\
CMS(7)   & 79.4/24   & 46.6$\pm$2.8 & 0.12 & $\!\!$0.15\\
ATLAS(7) & 33.6/33   & 28.6$\pm$1.5 & 0.15 & $\!\!$0.13\\
ALICE(13)& 46.0/44   & 35.4$\pm$1.4 & 0.14 & $\!\!$0.15\\
\hline
Eq. (\ref{eq_ATLAS})\\
\hline
ALICE(2.76) & \\
ALICE(7) & \multicolumn{8}{l}{\lw{almost the same results as the first above rows.}}\\
CMS(7)   & \\
ATLAS(7) & \\
ALICE(13) & 76.6/44 & 52.0$\pm$2.1 & 0.14 & $\!\!$0.15$\pm$0.00$\!\!$\\
  \hline
  \hline
Eq. (\ref{eq_BED}) & $\chi^2/$ndf & $A_p^{(M)}$ & $T$ (GeV) & $A_h$ & $p_{t0}$ (GeV) & $n_x$ & $\!\!\!\!\!$crossing$\!\!\!\!\!$ & $p_{t1}$(GeV) & $\!\!\!\beta_{\rm FD}\!\!\!$ (GeV$^{-1}$)\\
  \hline
ALICE(2.76) & 7.66/53 & 602$\pm$98 & 0.20$\pm$0.02 & 59$\pm$39 & 1.1$\pm$0.1 & 3.3 & 1.0 & 0.89$\pm$0.56 & 2.1 \\
ALICE(7) & 8.94/58 & 806$\pm$223 & 0.21$\pm$0.03 & 69$\pm$29 & 1.1$\pm$0.1 & 3.1 & 0.9 & 0.76$\pm$0.72 & 2.4 \\
CMS(7) & 7.21/20 & 9.20$\pm$2.77 & 0.21$\pm$0.02 & 1.99$\pm$0.56 & 1.4$\pm$0.1 & 3.3 & 0.8 & 10.0 & 0.91$\pm$0.01 \\
ATLAS(7) & 2.93/29 & 26.3$\pm$19.5 & 0.17$\pm$0.02 & 1.00$\pm$0.15 & 1.2 & 3.1 & 0.7 & 0.52$\pm$0.20 & 10.3$\pm$5.0 \\
ALICE(13) & 4.47/40 & 16.0$\pm$4.0 & 0.20$\pm$0.03 & 1.5$\pm$1.3 & 1.1$\pm$0.2 & 3.0 & 0.9 & 0.72$\pm$0.87 & 2.5 \\
\hline
Eq. (\ref{eq_Byl}) & $\chi^2/$ndf & $A_e$ & $T_e$ (GeV) & $A_h$ & $\!\!\!T_{\tiny BRR}$(GeV)$\!\!\!$ & $n$ & $\!\!\!\!\!$crossing$\!\!\!\!\!$\\
\hline
ALICE(2.76) & 7.93/55 &  558$\pm$32 & 0.15$\pm$0.01&  43$\pm$6                 & 0.65$\pm$0.02 & 3.3 & 0.7\\
ALICE(7) & 9.12/60 &  752$\pm$45.5 & 0.15$\pm$0.01 &        63$\pm$8           & 0.67$\pm$0.02 & 3.1 & 0.6\\
CMS(7)   & 70.8/22 & 8.84$\pm$1.29 & 0.20$\pm$0.01 & $\!\!$0.47$\pm$0.10$\!\!$ & 0.79$\pm$0.03 & 3.2 & 1.0\\
ATLAS(7) & 3.01/31 & 13.0$\pm$1.9  & 0.16$\pm$0.01 & $\!\!$0.98$\pm$0.13$\!\!$ & 0.69$\pm$0.02 & 3.1 & 0.7\\
ALICE(13)& 4.61/42 & 14.7$\pm$0.6  & 0.15$\pm$0.01 &        1.3$\pm$0.2        & 0.67$\pm$0.02 & 3.0 & 0.6\\
\hline
\end{tabular}
\end{table*}

As we are interested in analyses of $p_t$ distributions at 0.9 TeV, our results are shown in Table \ref{tab_09TeV}. As is seen in Tables \ref{tab_7TeV} and \ref{tab_09TeV}, estimated values by Eq. (\ref{eq_BED}) (the modified Hagedorn formula) and Eq.(\ref{eq_Byl}) (the BRR formula) are almost the same except for the temperatures: Those values of temperatures by Eq.(\ref{eq_Byl}) (BRR formula) are higher than those by Eq. (\ref{eq_BED}). The reason is probably attributed to the subtracted pion mass term in Eq. (\ref{eq_Byl}).
\begin{table*}[htbp]
  \caption{\label{tab_09TeV}Estimated values for the analysis of $p_t$ distribution at $\sqrt s =$ 0.9 TeV using Eqs. (\ref{eq_BED}) and (\ref{eq_Byl}) with $\eta_{\rm c} =0.8$ (ALICE), 2.4 (CMS), and 2.5 (ATLAS). The unit of the crossing is [GeV].}
  \begin{tabular}{lccccccccc}
  \hline
Eq. (\ref{eq_BED}) & $\chi^2/$ndf & $A_p^{(M)}$ & $T$ (GeV) & $A_h$ & $p_{t0}$ (GeV) & $n_x$ & $\!\!\!\!\!$crossing$\!\!\!\!\!$ & $p_{t1}$(GeV) & $\!\!\!\beta_{\rm FD}\!\!\!$ (GeV$^{-1}$)\\
  \hline
ALICE(0.9) & 7.45/47 & 448$\pm$88 & 0.19$\pm$0.02 & 27$\pm$25 & 1.2$\pm$0.2 & 3.6 & 1.0 & 0.72$\pm$0.99 & 2.0 \\
CMS(0.9) & 2.29/13 & 5.75$\pm$1.09 & 0.23$\pm$0.02 & 2.96$\pm$0.48 & 0.93$\pm$0.02 & 3.5 & 0.4 & 1.6$\pm$0.2 & 2.08$\pm$0.80 \\
ATLAS(0.9) & 15.1/24 & 6.38$\pm$0.55 & 0.22$\pm$0.01 & 2.8$\pm$0.2 & 1.00$\pm$0.01 & 3.7 & 0.4 & 1.61$\pm$0.08 & 1.80$\pm$0.20 \\
\hline
Eq. (\ref{eq_Byl}) & $\chi^2/$ndf & $A_e$ & $T_e$ (GeV) & $A_h$ & $\!\!\!T_{\tiny BRR}$(GeV)$\!\!\!$ & $n$ & $\!\!\!\!\!$crossing$\!\!\!\!\!$\\
\hline
ALICE & 7.64/49 &  395$\pm$22    & 0.16$\pm$0.01 &  25$\pm$5     & 0.65$\pm$0.03 &  3.6 & 0.7\\
CMS   & 3.58/15 &  9.46$\pm$1.86 & 0.15$\pm$0.01 & 0.63$\pm$0.15 & 0.63$\pm$0.03 &  3.6 & 0.7\\
ATLAS & 18.1/26 &  8.98$\pm$0.93 & 0.17$\pm$0.01 & 0.41$\pm$0.08 & 0.70$\pm$0.03 &  3.8 & 0.9\\
\hline
\end{tabular}
\end{table*}

\section{\label{sec4}Concluding remarks and discussion}

\paragraph{C1)} In 1983, Hagedorn proposed the formula shown as Eq. (\ref{eq_Hag}). The inverse power law should be changed according to concluding remark 2 (see C2) below). Moreover, the cross section measured by the rapidity ($y$) is assumed. However, because the pseudorapidity ($\eta$) is used in the measurement, the integral form is necessary, which is presented in Eq. (\ref{eq_BED}). In other words, Eq. (\ref{eq_BED}) can describe $\eta_c$ dependence in data on $d\sigma/2\pi p_t dp_t dy$. Moreover, we have assumed the Bose--Einstein distributions for describing the hadron gas. See Refs. \cite{Hagedorn:1983wk} and \cite{Landau:1965aa}.

\paragraph{C2)} To describe the data on the $p_t$ spectrum, including the smaller $p_t \le 1.0 - 2.0$ GeV region, the inverse power law with variable $p_t^2$ is better than that with a single $p_t$ (see the second term of Eq. (\ref{eq_BED}) and that of Eq. (\ref{eq_Byl})). This is observed in our analyses of the $p_t$ spectrum at $\sqrt s = 1.96$ TeV reported by the CDF collaboration.

\paragraph{C3)} The role of the kappa distribution in Eq. (\ref{eq_Byl}) is the same as that of the second term of Eq. (\ref{eq_BED}), i.e., $\theta$ function times the inverse power law distribution. See Eq. (\ref{eq_pt0}). Moreover, it can be emphasized that there is an interesting interrelation among Eqs. (\ref{eq_BED}), (\ref{eq_ATLAS}), and (\ref{eq_Byl}) as follows:
\begin{eqnarray*}
{\rm Eq.}\ (\ref{eq_BED}) \left\{
\begin{array}{l}
\mbox{Using the $\eta$ integration in Eq. (\ref{eq_BED}) \cite{Gradshteyn:2007}, and } \theta(x) = 0,\ 
I(\eta_c)\times f(n,\;T,\;p_t,\;m_{\pi})\mbox{ in Eq. (\ref{eq_ATLAS}) with }\medskip\\
I(\eta_c) = \dint_{-\eta_c}^{\eta_c} \!\!\! d\eta J(y(\eta))\\
\hspace*{9mm} = \left.\tanh^{-1}\left(\dfrac{p_t\sinh y(\eta)}{\sqrt{m_{\pi}^2+p_t^2\cosh^2 y(\eta)}}\right)\right|_{\eta = \eta_c} \longrightarrow \mbox{ Formula by ATLAS Coll. Eq. (\ref{eq_ATLAS})},\medskip\\
\\
\left. F_1\right|_{y(\eta)=0},\ x_{\pi} \to x_{\pi} - m_{\pi} \mbox{ and } \theta(x) = 1\ {\rm with\ Kappa\ dis.} \longrightarrow \mbox{ BRR formula Eq. (\ref{eq_Byl}). }
\end{array}
\right.
\end{eqnarray*}

\begin{figure}
  \centering
  \includegraphics[width=0.48\columnwidth]{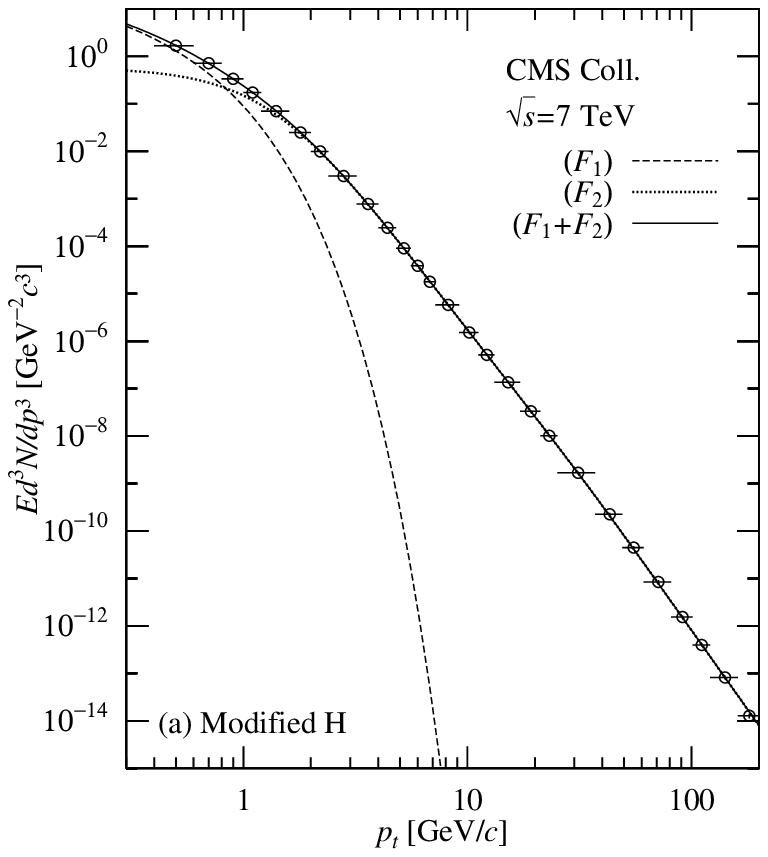}
  \includegraphics[width=0.48\columnwidth]{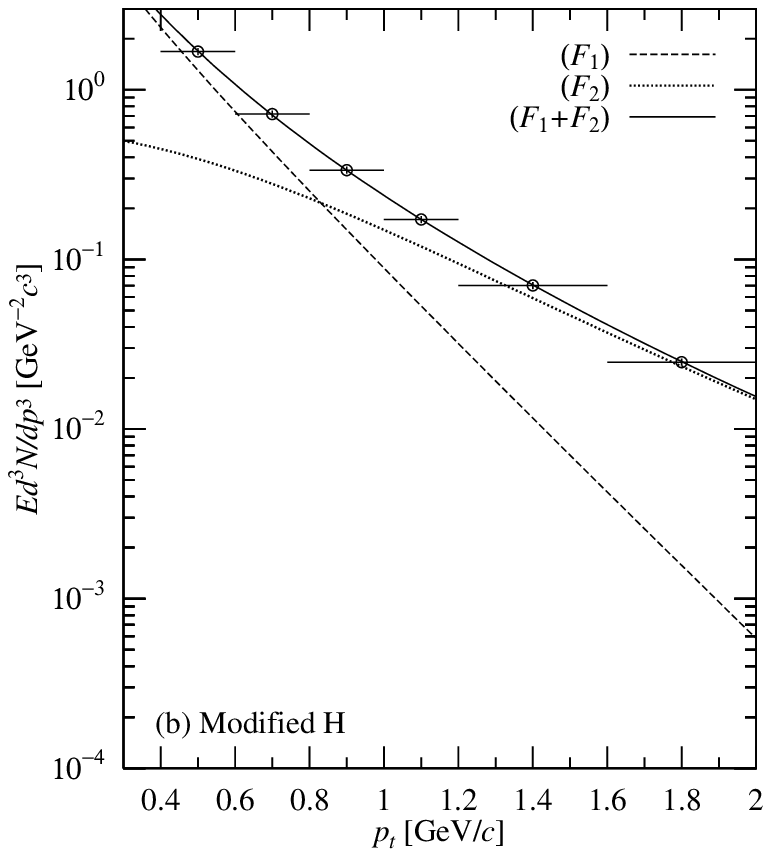}
  \caption{\label{fig_CMS}Analysis of $p_t$ distribution at $\sqrt s = 7$ TeV by CMS collaboration using Eq. (\ref{eq_BED}). $F_1$ and $F_2$ represent the results of the first and second terms of Eq. (\ref{eq_BED}), respectively. ($F_1 + F_2$) is the sum of these terms.}
\end{figure}

\paragraph{D1)} Concerning the different $\chi^2$ values obtained by Eqs. (\ref{eq_BED}) and (\ref{eq_Byl}) in analysis of data by CMS Collaboration at 7 TeV in Table \ref{tab_7TeV} (see Fig. \ref{fig_CMS}), we would like to point out that the range of $p_t$ is very wide ($0.4<p_t<201.2$ GeV). Second, we highlight that the estimated temperatures $T=0.21$ GeV in Eq. (\ref{eq_BED}) and $T_e=0.20$ GeV in Eq. (\ref{eq_Byl}) are almost the same. The latter is distinguished among results by Eq. (\ref{eq_Byl}). This phenomenon can probably be attributed to the lack of data between $0.2< p_t< 0.5$. Moreover, at present, the number of data points is 27. 

Thus, in a near future we will only be able to investigate whether Eq. (\ref{eq_BED}) works well when the number of data points increases so that it is close to the number of data points obtained by the CDF collaboration (230).

\paragraph{Acknowledgements:} One of authors (M. B.) would like to thank Department of Physics at Shinshu University for the hospitality.

\end{document}